\documentstyle[12pt,aasms4,psfig]{article}
\newcommand{\etal}{{et al. \,}}
\begin{document}

\title{A tidal extension in the Ursa Minor dwarf spheroidal galaxy}
\author{D. Mart\'\i nez-Delgado}
\affil{Instituto de Astrof\'\i sica de Canarias,
E38200  La Laguna, Tenerife, Canary Islands, Spain}
\authoremail{ddelgado@ll.iac.es}

\author{J. Alonso-Garc\'\i a}
\affil{Instituto de Astrof\'\i sica de Canarias,
E38200  La Laguna, Tenerife, Canary Islands, Spain}
\authoremail{jalonso@ll.iac.es}

\author{A. Aparicio}
\affil{Instituto de Astrof\'\i sica de Canarias 
E38200  La Laguna, Tenerife, Canary Islands, Spain\\
and Departamento de Astrof\'\i sica, Universidad de La Laguna}
\authoremail{aaj@ll.iac.es}

\author{M. A. G\'omez-Flechoso}
\affil{Geneva Observatory, CH-1290 Sauverny, Switzerland}
\authoremail{ddelgado@ll.iac.es}

\begin{abstract}

We report the detection of main-sequence and blue horizontal-branch stars of the Ursa Minor dwarf spheroidal galaxy beyond its tidal radius, indicating the existence of a possible tidal extension in this satellite of the Milky Way. This tidal extension could spread out well beyond the area covered in our survey ($R>80\arcmin$),
as suggested by the presence of a ``break'' to a shallower slope observed in its density profile. The  $V$-band surface brightness for this possible tidal extension range from 29.8 to 31.5 mag arcsec$^{-2}$. The area covered in our survey ($\sim$ 1.65 
deg$^{2}$) is not enough to discriminate if this extra-tidal
population is  part of a tidal tail or an extended halo around the galaxy. 

The existence of this tidal extension in Ursa Minor indicates that this satellite is currently undergoing a tidal disruption process by the Milky Way. We discuss the possibility of a  tidal origin for the  high mass-to-light ratio observed in this galaxy on the basis on our result and  recent theoretical simulations of the tidal disruption of dwarf satellites in the Galactic halo.

\end{abstract}

\section{Introduction}

In hierarchical clustering scenarios for galaxy formation, such as cold dark matter-dominated cosmologies (White \& Rees 1978; Blumenthal et al.  1984; Dekel \& Silk 1986), dwarf galaxies should have formed prior to the epoch of
giant galaxy formation and would be the building blocks of larger galaxies.
The picture of building the Galactic halo from merging ``fragments'', which Searle \& Zinn (1978, hereafter SZ) 
proposed on the basis of the properties of the Milky Way
globular clusters, is regarded as the local manifestation of this 
galaxy formation scenario.

Recent evidence shows that the inner and intermediate region of the Galactic halo
has mostly been formed in a fast process (Rosenberg et al.  1999). However, several results indicate that at least a part of the halo originated in a process
similar to the SZ scenario and compatible with the hierarchical galaxy formation theory. The discovery of the Sagittarius
dwarf galaxy (Ibata, Gilmore, \& Irwin 1994), in the process of
dissolving into the Galactic halo, favored the hypothesis
that merging
events can take place in the Milky Way, whose  full formation
history (through satellites merging into it) might not have  finished
yet. Isolated stellar tidal streams have also been identified in the
Galactic halo by many authors (Majewski
1998; C$\hat{o}$t\'e et al.  1993; Arnold \& Gilmore 1992; Helmi et al. 1999). Lynden-Bell \& Lynden-Bell (1995) found that
several dwarf spheroidals (dSphs) and outer globular clusters appear to lie along two
distinct streams that may be the remnants of larger parent
satellite galaxies or SZ ``fragments''. Some observational evidence also exists
on the presence of tidal tails in  dSph galaxy companions of the Milky Way
(Kuhn, Smith, \& Hawley 1996; Irwin \& Hatzidimitriou 1995, hereafter
 IH; Majewski et al.  2000; Piatek \etal 2001). The possibility that accretion
events may leave observable fossil records in the halo is also
supported by theoretical models of tidally disrupted dSph satellites
(Johnston, Spergel, \& Hernquist 1995; Oh, Lin, \& Aarseth 1995; Piatek \& Pryor 1995; G\'omez-Flechoso, Fux, \& Martinet 1999).

Despite the evidence mentioned above, the question of whether other Milky Way
satellites could have been tidally disrupted by the Milky Way is still open
to debate. On the basis of this discussion, it is very
important to investigate whether Galactic  dSph
satellites display tidal tails to understand if merging events played an important
(Mateo 1996) or minor role (Unavane, Wise, \& Gilmore 1996) in the formation of
the Galactic halo. Moreover, determining if and how dSphs dissolve is
fundamental to several open questions concerning the dark matter
content and the age spread of the halo, because the majority of them
exhibit complex star formation histories (Mateo 1998) and apparently
significant amounts of dark matter (Aaronson 1983; Olszewski, Pryor, \& Armandroff 1996). These issues are intimately
related to the difficulties in the interpretation of microlensing
towards the LMC (Zhao 1998; Alcock et al.  1997; Zaritsky \& Lin
1997) and to the existence of intermediate-age stars (Preston, Beers, \& Schectman 1994) and main sequence A stars (Rodgers, Harding, \& Sadler 1981; Yanni et al.  2000) above the
Galactic plane. The availability of a new generation of wide-field CCD
cameras offers for the first time a good opportunity of successfully addressing these issues.

	Ursa Minor (UMi) is one of the closest satellites of the Milky
Way ($d=69$ kpc) and is a strong candidate for being in the terminal phase of its complete tidal disruption (Hodge \& Michie 1969). It also has the largest ellipticity (0.55; IH) if we exclude Sagittarius and displays, together with Draco, the highest observed velocity dispersion (Armandroff, Olszewski, \&  Pryor 1995). It also
belongs to one of the intriguing alignments of dSphs on the sky found by
Lynden-Bell \& Lynden-Bell (1995), suggesting that it could have been stripped from a larger ``building-block'' long time ago. This makes its study quite interesting in the aforementioned context. In this letter, we report the detection of a tidal extension in  UMi dSph from a systematic wide field survey using broad-band deep photometry.

\section{OBSERVATIONS AND DATA REDUCTION}

UMi dSph was observed in $B$ and $R$ Johnson--Cousins filters with the Wide
Field Camera (WFC) at the prime focus of the 2.5 m Isaac Newton
Telescope (INT) at the Roque de los
Muchachos Observatory on the island of La Palma (Canary Islands, Spain). The
WFC holds four 4096 $\times$ 2048 pixel EEV CCDs with pixel size
$0\farcs33$, which provides a total field of about $35\times
35$arcmin$^2$. Three fields were observed covering a total area of about
1 deg$^2$ as shown in Figure 1 (see Mart\'\i nez-Delgado \& Aparicio 1999
for more information). Total integration times were 1800 s in both filters.

Bias and flatfield corrections were done with IRAF. DAOPHOT and ALLSTAR
(Stetson 1994)  were then used to obtain the instrumental photometry of the
stars. About 45~000 stars were measured in both bands. An overlap region between the adjacent fields of UMi was use to obtain a common internal photometric system for the galaxy. Atmospheric extinction
and transformations to the standard Johnson--Cousins photometric system were
obtained from observation of several standard stars of the Landolt (1992)
list. Details about the photometric transformation are given in Mart\'\i
nez-Delgado et al. (2001). Suffice it to say here that measurements of 17
standards were used, and that the extinction was determined with an accuracy
better than 0.02 mag, while the photometric transformation zero-point errors
are 0.008 mag in $B$ and 0.013 in $R$. The photometric transformations between the different WFC chips were obtained from observations of several standard fields in each different chip. These photometric zero-points between different 
chips were estimated with an accuracy better than 0.01. mag. The total zero-point errors of the photometry are therefore about 0.025 in both filters. 

\section{METHODOLOGY }

The detection of tidal tails in Local Group dSphs is very challenging due to
their large angular sizes and low surface brightnesses (LSB), and requires
using wide-field observations and a careful analysis of the foreground and
background contamination (since extended objects are rejected by DAOPHOT and
ALLFRAME the background contamination will presumably include only
stellar-shaped objects). An efficient technique is based on the analysis of
wide-field, deep color--magnitude diagrams (CMDs; e.g., Mateo,
Olszewski, \& Morrison 1998). The LSB tail can be detected through star counts
at the old population main-sequence (MS) turn-off region. This feature is the
most densely populated in the CMD and thus provides the best contrast against
the foreground and background population. In the case of UMi, its extra-tidal
structure can be also traced by the presence of blue horizontal branch (BHB)
or blue straggler (BS) stars, because a gap exists in the distribution of
contaminating foreground and background objects in coincidence with the BHB
and BS regions.

Determining whether an extended population in UMi is or not a tidal stream is
also challenged because of the uncertainties about its angular size. The
early work by Hodge (1964) provides a tidal radius of $r_{\rm t}=75'\pm 25'$, but
model fitting to the same data by Lake (1990) and by Pryor \& Kormendy (1990)
yield $r_{\rm t}=59'$ and $r_{\rm t}=31\farcm3$ respectively. IH
 made a new study of the structural parameters of UMi using a 
wide-field scanned photographic plate. They found a tidal radius of
$r_{\rm t}=50\farcm2$, reporting also the presence of possible extra-tidal
stars. However, Kleyna \etal (1998), from two-color, CCD photometry, derived
$r_{\rm t}=34 \arcmin$ and argued that the larger value reported by the former
authors might be the result of a biased estimate of the background
contamination, which would also result in the detection of fake extra-tidal
stars. Indeed, the two-color photometry of Kleyna \etal (1998) allows
the removal of many contaminating objects from the density distribution on the
basis of their place in the CMD. In this paper, we will adopt the Kleyna
\etal tidal radius, although we will also check the presence of extra-tidal
stars beyond the IH value.

\section{THE TIDAL EXTENSION OF THE UMI DWARF SPHEROIDAL}

To study the extension of UMi we have divided the galaxy into
concentric elliptical annuli of the same eccentricity and position angle,
for which we have adopted the IH values, and
increasing semi-major axis (see Figure  1). 

Figure  2 shows the $[(B-R),V]$ CMDs of the stars in three selected elliptical
annuli and of a control field situated $3 \arcdeg$ south of the center of
UMi. The selected annuli are a) the central region ($R<6\farcm6$; with $R$
the ellipse semi-major axis); b) the region between the Kleyna et al. and IH
$r_{\rm t}$ ($34.0 \arcmin < R < 50.6
\arcmin $); and c) the region between the IH $r_{\rm t}$ and the limit of our data ($ 50.6 \arcmin < R < 78.1 \arcmin $).

Visual inspection of the CMDs indicates that the galaxy extends beyond the
$r_{\rm t}$ of Kleyna et al. (Figure  2b), as indicate the presence of
BHB and BS stars and other clear CMD features such as the MS turn-off and the
sub-giant sequence. A few BHB stars are also observed beyond the IH 
$r_{\rm t}$ (Figure  2c), meaning the presence of the galaxy's
stars even in this outer region. This is confirmed by the detection of a
clump at the position of the MS turn-off ($(B-R)\sim$ 0.8, $V$ = 23.5), 
which is
not observed in the control field.

We use the star counts method mentioned in \S 3 (see also Mateo \etal  1998) to  estimate quantitatively the extension and surface brightness of this possible tidal extension in UMi. We only counted stars within the MS box  shown in Figure  2a, subtracting the counts from the control field for statistically removing the contribution of non-UMi stars in the box.  We find that the differential reddening between the UMi and control fields is negligible from the extinction maps by Schlegel, Finkbeiner \& Davis (1998). Figure  3 (upper panel) shows the logarithm of the raw MS stars counts per unit area (in arcmin$^{2}$) in each ellipse. The solid line shows the mean value for the control field with its Poisson error represented as dot--dashed lines. The logarithm of the net stars per unit area (once  the contaminating
objects are removed) are shown  in the lower panel, where the error bars denote the Poisson uncertainties of each point. 

Interestingly, the raw surface distribution shows a plateau for $55\arcmin <a <
78.1$, which could be interpreted as the background density having been reached.
This would mean that our density value obtained from the control field is too low. To check this possibility, we have estimated the number
of galaxy members expected in the annuli $ 50.6 \arcmin < R < 78.1 \arcmin $
(Figure 2c) adopting the mean value of the background level from the radial profile at large radii (R$>60 \arcmin$) in Figure 3a (--0.04 in units as Figure 3a). The result is that only $\sim$ 75 stars are expected in the MS box for the region of this annuli covered in our survey (see Figure 1), and therefore no traces of the galaxy would be observed in the CMD plotted in Figure 2c. However, the presence of BHB and MS stars in this CMD shows that this is not the case. This suggests that this change to a shallower slope in the density profile could be a signature of the presence of a very faint tidal tail that extends to larger radii in this galaxy. Further observations are necessary to confirm this.

The surface brightness of this tidal extension can be estimated from our net star counts plotted in Figure 3. With this purpose, we normalize to the $V$-band surface brightness (SB) estimated in the center region of UMi ($\Sigma$ = 25.5 mag/$arcsec^{2}$) by Mateo (1998) to express the star counts in SB units. This is also shown in the lower panel of Figure 3. The inferred $V$-band SB profile for the possible tidal extension ranges from 29.8 to 31.5 mag arcsec$^{-2}$.

\section{ DISCUSSION}

We report the detection of stellar members of the UMi dSph beyond the $r_{\rm t}$ given in previous studies, indicating the existence of a possible tidal extension in this galaxy. This tidal extension
could spread out well beyond the area covered in our survey ($R>80\arcmin$),
as suggested by the presence of a ``break'' to a shallower slope observed in its density profile (see Johnston, Sigurdsson \& Hernquist 1999). Unfortunately, our data are insufficient to conclude 
whether this extra-tidal
population is a part of a tidal tail or an extended halo around the galaxy. 

The existence of this tidal extension in UMi indicates that this satellite
is undergoing a tidal disruption process. In this context, it is
important to discuss the possibility of a possible tidal
origin for the  UMi's   observed high radial-velocity dispersion, as 
 has been suggested by several authors (Kuhn \& Miller 1989; G\'omez-Flechoso et al.  1999; Kroupa 1997). However, the interpretation of our results is difficult due to the open theoretical controversy about the origin of the high mass-to-light ($M/L$) ratios in dSphs. In the case of a dSph satellite significantly perturbed by tides, 
one of the main alternative explanations for these large
$M/L$ values is that the assumption of virial equilibrium may be not
fulfilled. In these circumstances, a system having a low mass-to-luminosity ratio can present a large velocity dispersion (Kuhn \& Miller 1989; G\'omez-Flechoso et al.  1999) due to
the tides increasing the energy of the stars in
the dwarf galaxy and, accordingly, their internal velocity dispersion. However, Piatek \& Pryor (1995) and  Oh  et al. (1995) found that, without dark matter, the velocity dispersion could not
be as high as is observed, even considering its tidal disruption.
Another alternative explanation has been proposed by Kroupa (1997) and Klessen \& Kroupa (1998), who suggested that these dSphs with large apparent ($M/L$) are actually long-lived tidal remnants whose main axis, together with the main axis of the
velocity ellipsoid, could eventually be oriented close to the line of sight. In
such a case, an observer would derive values for $M/L$ that are much larger than
the true $M/L$ ratio of the stars and would not need large quantities of
dark matter to be accounted for. However, the observed width of the HB in
Draco (Aparicio, Carrera, \& Mart\'\i nez-Delgado 2000) and UMi (Mart\'\i nez-Delgado et al.  2000) seems to indicate that this is not the case for these galaxies.

  The presence of substructure in the main body of UMi also strengthens
 the idea that
this satellite is being destroyed by  Milky Way tides. Olszewski \& Aaronson (1985) reported that the surface stellar density of UMi
is patchy, finding two regions of high stellar density separated by a valley
close to the center of the galaxy. This lumpy structure was also found in the recent studies by Irwin \& Hatzidimitriou (1995) and Kleyna et al. (1998). We confirm the presence of lumpiness and asymmetry in the stellar distribution of UMi along its major axis (Mart\'\i nez-Delgado et al.  2000), although we are currently carrying out an analysis to test its statistical significance.  If this substructure is real, more 
elaborate models, including details of the substructure and the presence
of tides, will be needed to estimate the real dark-matter content of UMi.
A low dark-matter content allows the  formation of tidal tails, besides the presence of substructures formed by tidal interaction. However, tidal tails and substructures are more difficult to understand in the context of a massive dark halo. A more extended systematic survey of the outer regions of Ursa Minor and accurate velocity measurements of these extra-tidal component are required to understand this important question concerning the dark matter content in UMi.

This work is based on observations made with the 2.5 m Isaac Newton Telescope operated on the island of La Palma by the Isaac Newton Group in the Spanish Observatorio del
Roque de Los Muchachos of the Instituto de Astrof\'\i sica de Canarias.

\newpage

\begin{figure}
\centerline{\psfig{figure=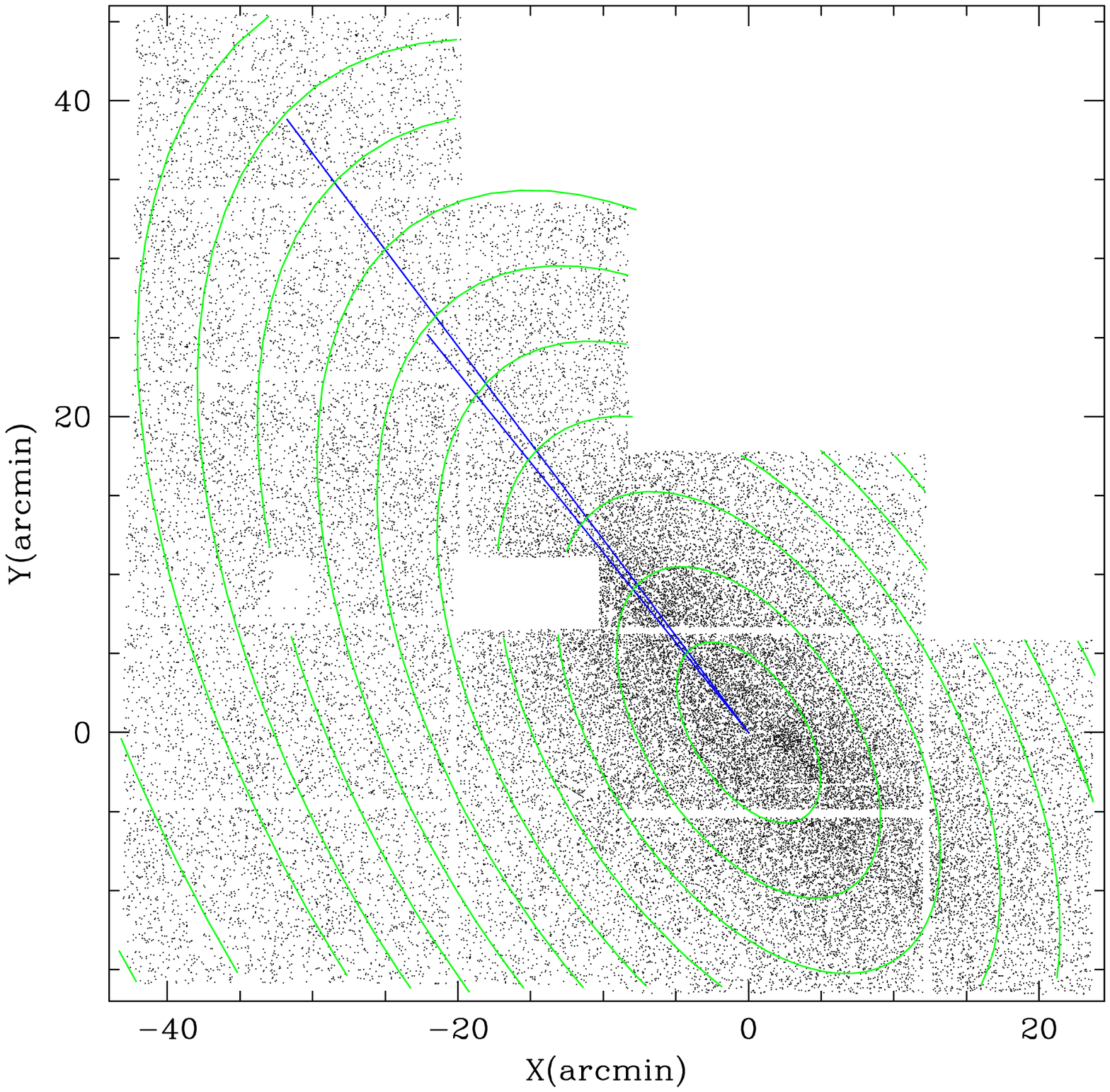,width=20cm}}
\figcaption[fig2.eps]{ The spatial distribution of the stars measured 
in our survey,
together with the ellipses used to divide the galaxy for the density-profile study. The first ellipse has a semi-major axis of 6.6$\arcmin$. Further ellipses are obtained with a semi-major axis step of 5.5$\arcmin$. The 
50.6$\arcmin$ and 34$\arcmin$ tidal
radii obtained by Irwin \& Hatzidimitriou (1995) and Kleyna et al. (1995),
respectively, are indicated by the arrows.
\label{ima_1}}
\end{figure}

\newpage

\begin{figure}
\centerline{\psfig{figure=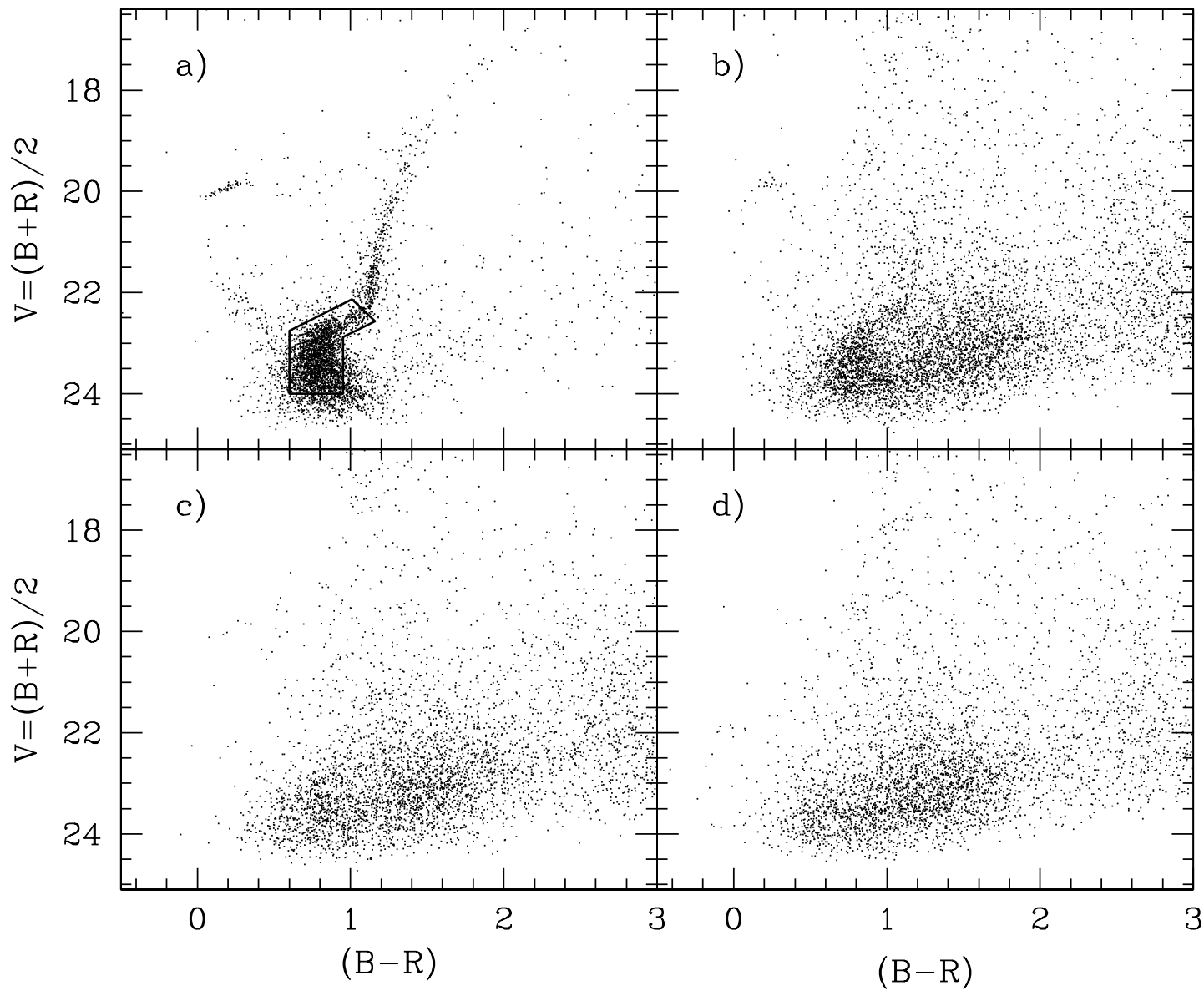,width=20cm}}
\figcaption[fig1.eps]{CMDs of the UMi fields. Panel a) shows the CMD of
the central region ($ R < 6.6\arcmin $), with the box used to count stars in the MS region.
Panels b) and c) show two extra-tidal fields of
the galaxy ($ 34.0 \arcmin < R < 50.6 \arcmin $ and $ 50.6 \arcmin < R < 78.1 \arcmin $ respectively; see \S 3). The detection of BHB and MS stars in these
two fields reveals the existence of a tidal extension in UMi. Panel d) shows
the CMD of a control field, $\sim 3\arcdeg$ South from the center of the
galaxy. \label{ima_1}}
\end{figure}

\newpage

\begin{figure}
\centerline{\psfig{figure=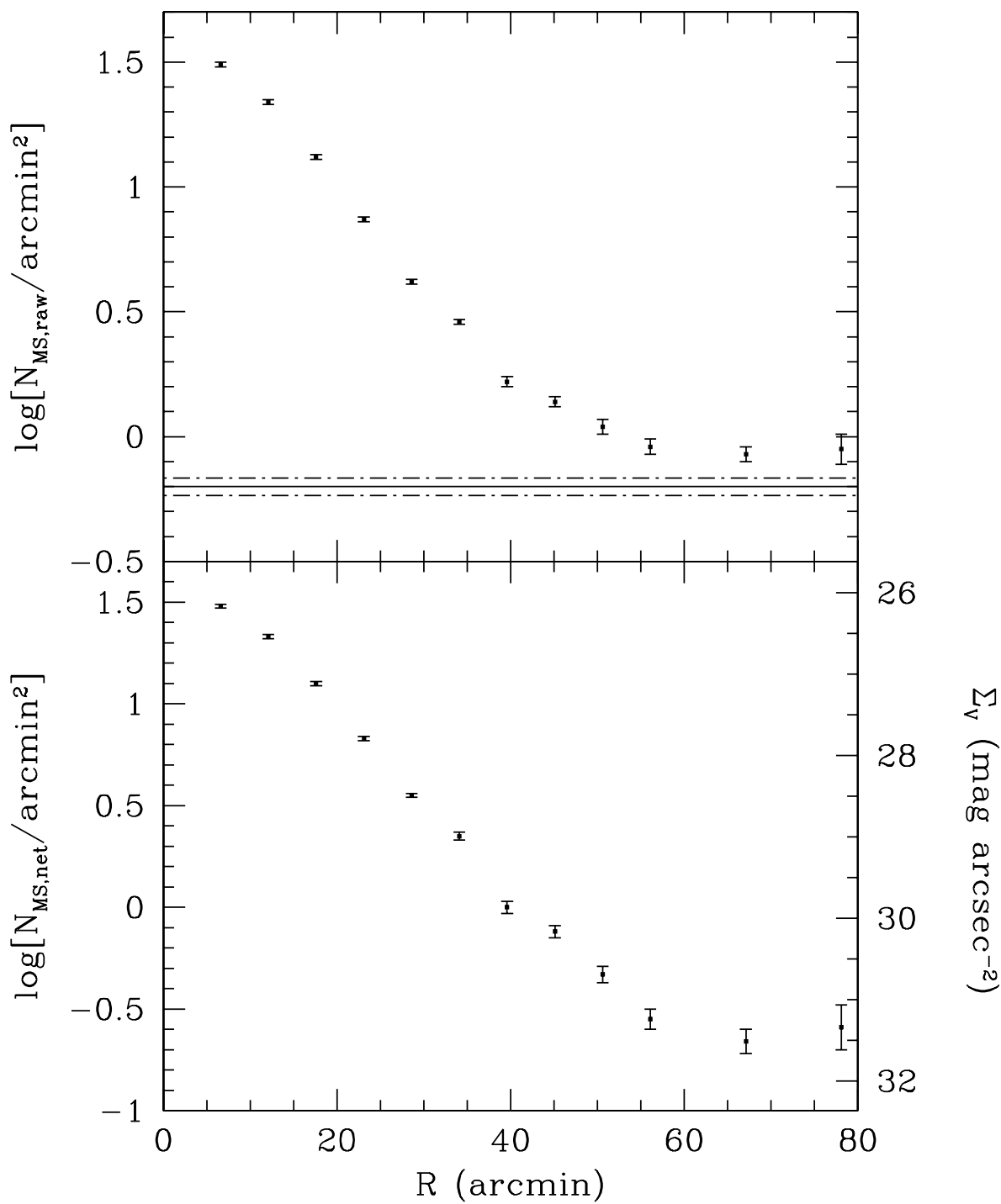,width=20cm}}
\figcaption[ima_1.eps]{Upper panel: the raw star counts per unit area in the
MS box (see Figure 1a) in each Umi ellipse plotted in Figure 2. The solid line shows
the foreground level estimated from the control field. Dot--dashed lines show the
 Poisson uncertainties. Lower panel: the net star counts per unit area  after
subtraction of the foreground density.
\label{ima_1}}
\end{figure}

\end{document}